\documentclass[iop]{emulateapj}
\usepackage{amsmath}
\usepackage{apjfonts}
\usepackage{natbib,graphicx}
\usepackage{longtable}
\newcommand{\cp}{\citep}
\newcommand{\ct}{\citet}

\begin{document}

\title{Re-inflated Warm Jupiters Around Red Giants}
\author{Eric D. Lopez}
\affil{Institute for Astronomy, Royal Observatory Edinburgh, University of Edinburgh, Blackford Hill, Edinburgh, UK} 
\author{Jonathan J. Fortney}
\affil{Department of Astronomy and Astrophysics, University of California, Santa Cruz, CA 95064} 

\begin{abstract}
Since the discovery of the first transiting hot Jupiters, models have sought to explain the anomalously large radii of highly irradiated gas giants. We now know that the size of hot Jupiter radius anomalies scales strongly with a planet's level of irradiation and numerous models like tidal heating, ohmic dissipation, and thermal tides have since been developed to help explain these inflated radii. In general however, these models can be grouped into two broad categories: 1) models that directly inflate planetary radii by depositing a fraction of the incident irradiation into the interior and 2) models that simply slow a planet's radiative cooling allowing it to retain more heat from formation and thereby delay contraction. Here we present a new test to distinguish between these two classes of models. Gas giants orbiting at moderate orbital periods around post main sequence stars will experience enormous increases their irradiation as their host stars move up the sub-giant and red-giant branches. If hot Jupiter inflation works by depositing irradiation into the planet's deep interiors then planetary radii should increase in response to the increased irradiation. This means that otherwise non-inflated gas giants at moderate orbital periods >10 days can re-inflate as their host stars evolve. Here we explore the circumstances that can lead to the creation of these "re-inflated" gas giants and examine how the existence or absence of such planets can be used to place unique constraints of the physics of the hot Jupiter inflation mechanism. Finally, we explore the prospects for detecting this potentially important undiscovered population of planets.
\end{abstract}

\section{Introduction}

In 1999 the first transiting extrasolar planet was discovered by \ct{Charbonneau2000} and \ct{Henry2000}. While in many ways this represented an enormous triumph for both exoplanet observations and theory, it also presented a key mystery as the planet's measured radius was much larger than expected \cp[e.g.,][]{Bodenheimer2001,Guillot2002}. Ever since, theorists have been challenged to explain the anomalously large radii of this and many other hot Jupiters. Planet structure and evolution models had predicted that there should be a clear maximum size for evolved giant planets \cp[e.g.,][]{Burrows1997}. In the absence of any additional energy source even a completely gaseous planet should never be larger than $\approx$ 1.2 Jupiter Radii by the time it is several billion years old \cp[e.g.,][]{Fortney2007}. Nonetheless, at short orbital periods there are now dozens of highly inflated hot Jupiters, some with radii that are almost $\sim$2 R$_{\mathrm{J}}$ \cp[e.g.,][]{Hebb2009,Anderson2010}. Although extremely young planets can have radii this large, due to their initial heat from formation, this cannot explain the observed hot Jupiter population, which generally orbit stars several Gyrs old.

\begin{figure}
  \begin{center}
    \includegraphics[width=3.5in,height=2.5in]{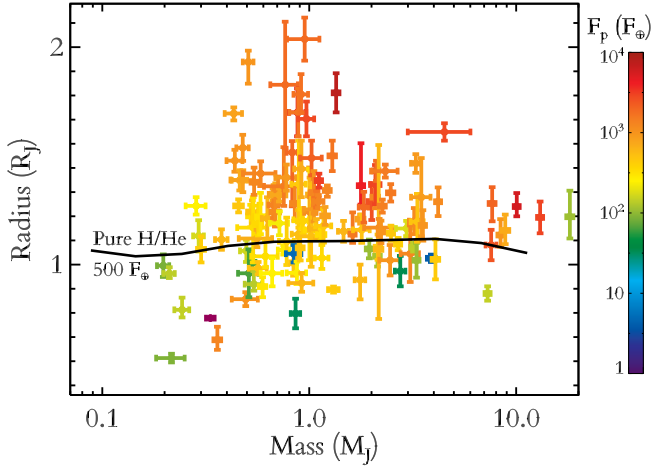}
  \end{center}
  \caption{Planetary masses and radii in Jupiter masses and radii for all transiting planets with measured masses >100 M$_{\mathrm{\oplus}}$ according to exoplanets.org \cp{Wright2011}. Each planet is color-coded by the current level of irradiation that it receives from its parent star, relative to F$_{\mathrm{\oplus}}$ the level of irradiation that the Earth receives from the Sun. The black line shows the mass-radius relationship for core-less gas giants at solar metallicity, 5 Gyr, and 500 F$_{\mathrm{\oplus}}$ taken from \ct{Fortney2007}. This indicates the maximum possible size of hot Jupiters in the absence of the unknown inflationary mechanism. \label{mrfig}}
\end{figure}

\begin{figure}
  \begin{center}
    \includegraphics[width=3.5in,height=2.5in]{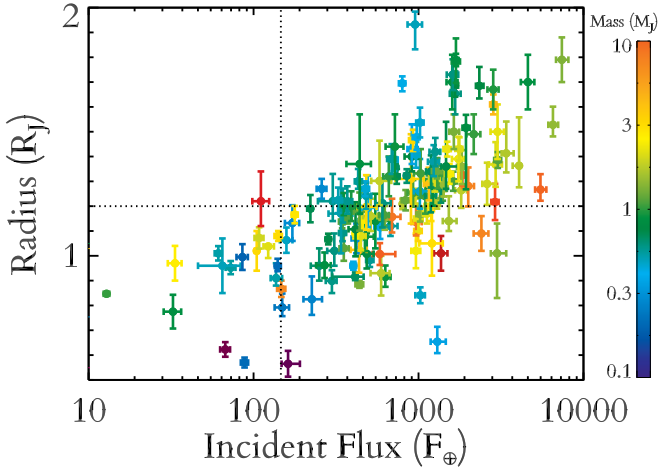}
  \end{center}
  \caption{Planetary radius in Jupiter radii vs. insolation, relative to the insolation of the Earth, for all transiting planets with measured masses >100 M$_{\mathrm{\oplus}}$ according to exoplanets.org \cp{Wright2011}. Each planet is color-coded by its mass in Jupiter masses. This shows the clear correlation between planetary radii and irradiation at fixed planetary mass. The horizontal dotted line shows the approximate threshold for planets to be defined as inflated, while the vertical dotted line shows the approximate irradiation threshold from \ct{Miller2011} below which there are no clearly inflated planets. \label{inffig}}
\end{figure}

Moreover, not all hot Jupiters are inflated to the same degree; instead, we see that even among the highly irradiated gas giants, the most irradiated planets are much more inflated than those that are slightly further out \cp[e.g.,][]{Laughlin2011}. Figure \ref{mrfig} shows the masses and radii of all transiting planets with measured masses >100 M$_{\mathrm{\oplus}}$ according to exoplanets.org \cp{Wright2011}. In addition, we have color-coded each planet by the current level of irradiation that it receives from its parent star, relative to F$_{\mathrm{\oplus}}$ the level of irradiation that the Earth receives from the Sun today. Similarly, in Figure \ref{inffig} we have re-plotted the radii of these planets as a function of their irradiation, color-coding instead by planet mass, in order to further highlight the connection between planetary irradiation and inflated radii. From this pair of figures we can see that the most inflated planets with radii $\gtrsim$1.5 R$_{\mathrm{J}}$ (e.g, WASP-12b, WASP-17b, HAT-32b, etc.) are those with the highest incident fluxes $\gtrsim$ 1000 F$_{\mathrm{\oplus}}$ and relatively low masses $\lesssim$ 1 M$_{\mathrm{J}}$. Less irradiated planets are at most modestly inflated. Likewise, planets with masses $\gtrsim$3 M$_{\mathrm{J}}$ are difficult to inflate, due to their larger gravitational binding energies. On the other hand, the lack of highly irradiated inflated with masses $\lesssim$0.5 M$_{\mathrm{J}}$ may be due to the destruction of such planets by Roche-lobe overflow and tidal in spiral \cp[e.g.,][]{Jackson2010}.

Whatever the nature of the inflation mechanism, clearly the degree of hot Jupiter inflation is in some way intimately tied to a planet's orbital period or level of irradiation. For the bulk of modestly of inflated hot Jupiters $\sim$1.5 R$_{\mathrm{J}}$, their radii can be explained if $\sim$1\% of their incident irradiation is somehow deposited into their deep convective interiors \cp{Batygin2010}. However, the large diversity in inflated radii in points to a large variation in the mechanism that causes radius inflation. As pointed out by \ct{Socrates2013}, the range of internal heating rates needed to maintain the observed range of inflated radii spans 4-5 orders of magnitude. 

In the past few years, many models have been put forward to explain this unknown inflation mechanism \cp{Spiegel2013}. Initially, heating due to tidal circularization looked like a promising explanation \cp[e.g.,][]{Bodenheimer2001}. At longer orbital periods there is a large population of giant planets on eccentric orbits \cp[e.g.,][]{Ford2008}. Moreover, since giant planets cannot form at highly irradiated orbits \cp[e.g.,][]{Alibert2005,Ida2008}, it has long been suggested that hot Jupiters may have arrived at their current locations on highly eccentric orbits due to planet-planet scattering \cp[e.g.,][]{Ford2008} or Kozai oscillations \cp[e.g.,][]{Fabrycky2007}. Those orbits would then circularize due to tidal dissipation and in the process deposit an enormous amount of heat in the planets interior. However, since the tidal heating rate goes as the inverse of the circularization timescale, the tidal heating rate is only large when planets are in the process of rapidly circularizing, it should be relatively unusual to find old inflated planets on circular orbits, since to explain their inflated radii circularization must have been quite recent\cp[e.g.,][]{Miller2009}. Self-consistent calculations of heating due to tidal circularization predict that only a fraction of hot Jupiters should be tidally inflated \cp{Leconte2010}. This predicted rarity stands in sharp contrast to the large population of inflated hot Jupiters shown in Figure \ref{mrfig}, meaning that tidal circularization cannot be the main culprit.

More recently, \ct{Arras2009,Arras2010} and \ct{Socrates2013} put a new spin on the possibility of tidal heating. \ct{Arras2009} pointed out that in addition to the normal gravitational tidal bulge, highly irradiated gaseous planets should have an additional thermal tidal bulge. Near the planet's sub-stellar point surface temperatures will be at their highest, therefore according the ideal gas law, at a given pressure level, density will be at its lowest there. For a given pressure level, there will be a slight mass deficit at the sub-stellar longitude and slight mass enhancements near the terminators. This thermal tidal bulge contrasts with the gravitational tidal bulge, which will be oriented towards the parent star. Moreover, due to the planet atmosphere's thermal inertia, the thermal bulge will not be oriented exactly 90 degrees away from the gravitational tidal bulge. This means that there will be a non-zero cross-product between the two bulges allowing them to torque on each other and slightly spin up the planet's rotation. This in turn means that a planet will no longer be in a spin synchronous orbit, leading to tidal heating of its interior. For highly irradiated planets, the mass in the thermal tidal bulge becomes non-trivial, potentially allowing thermal tides to produce the necessary heating rates to explain inflated hot Jupiters. Moreover, since the amplitudes of the thermal and gravitational tides both depend on distance from the host star, their product (and therefor the thermal tidal heating rate) will be highly sensitive to the level of irradiation, allowing for the large dynamic range of heating rates needed to explain the observed population \cp{Socrates2013}.

Alternatively, some models suggest that there is no need to actively heat a planet's interior. Instead, if a mechanism can sufficiently slow the radiative cooling from a planet's atmosphere then it may be possible for a hot Jupiter to retain much more of its heat from formation, allowing it to maintain a large radius for billions of years. \ct{Burrows2007} showed that some of the moderately inflated planets could be explained by delayed cooling due to enhanced atmospheric opacity. Similarly, \ct{Baraffe2008} showed that strong interior composition gradients could inhibit convection and delay cooling. 

One particularly popular explanation of inflated hot Jupiters is heating to ohmic dissipation \cp[e.g.,][]{Batygin2010,Perna2010,Batygin2011,Huang2012,Menou2012,Wu2013}. At temperatures $\gtrsim$ 1500 K, corresponding to incident fluxes $\gtrsim$ 900 F$_{\mathrm{\oplus}}$, alkali metals in exoplanetary atmospheres will thermally ionize \cp{Batygin2010}. When combined with the rapid wind speeds characteristic of hot Jupiters \cp[e.g.,][]{Showman2009}, these ions create significant currents in planetary atmospheres. These atmospheric currents in turn induce response currents in a planet's interior, where hydrogen is in the highly conductive liquid metallic phase. These interior currents are then dissipated due to the large resistivity at the molecular/metallic phase boundary, potentially heating planetary interiors and inflating radii. 

However, the degree of interior heating and radius inflation in ohmic heating models depends strongly on the depth where the dissipation takes place. If the dissipation takes place at pressure  $\gtrsim$1 kbar, then the energy will be deposited into the interior convective zone, effectively heating the entire planet interior \cp[e.g.,][]{Batygin2010}. On the other hand, if the heating is deposited at pressures $\lesssim$ 100 bar, only the radiative atmosphere will be heated \cp[e.g.,][]{Huang2012,Wu2013}. This will push the radiative convective boundary deeper into the planet leading to slower cooling, but will not actively heat the interior \cp{Huang2012}. Thus ohmic heating can act either to actively heat the interior or merely delay cooling depending on the depth and extent of the dissipation layer, parameters which are both difficult to determine and model dependent \cp{Spiegel2013}.

\subsection{Re-inflation: a New Test of Hot Jupiter Inflation Models}

For our purposes here, we classify inflationary models into two broad categories. In class I models, a fraction of the stellar irradiation incident on a planet is deposited into the planet's deep interior, directly heating the planet's interior adiabat and inflating its radius. Conversely, in class II models no energy is deposited into the interior, instead the inflationary mechanism simply acts to slow radiatively cooling through the atmosphere, allowing a planet to retain more of its heat from formation. While both classes of models may be capable of explaining the currently know population of hot Jupiters, here we propose a new way to distinguish between inflationary models.

One possible test for models of hot Jupiter inflation is to examine their response to large changes in the irradiation a planet receives from its parent star, such as happens when their parent star leaves the main sequence and moves up the sub-giant and red giant branches. In the process the parent star's luminosity and therefore the irradiation on a planet increases by several orders of magnitude, eventually making even long-period giants into hot Jupiters \cp{Spiegel2012}. However, unlike the long period Jupiter analogs studied in \ct{Spiegel2012} planets with periods $\sim$20 days will experience high levels of irradiation for more than 100 Myr. Consequently, If the inflation mechanism is a class I model that effectively deposits incident irradiation into the interior, then the planet's radius should respond to this dramatic increase in irradiation. On the other hand, in if the mechanism is a class II model that merely delays but does not reverse cooling, then there should be no effect on planet radius.

There are two significant hurdles for this proposal. The first issue is a question of timescales. The post-main sequence lifetime of sun-like or more massive stars is relatively short. Moreover, as the star expands it will eventually engulf any short to moderate period planets. Can any heating mechanism deposit sufficient energy in a planet's interior to produce inflated radii before that planet is swallowed by its host star? \ct{Spiegel2012} examined this possibility for true Jupiter analogs and found that the RGB evolution was too rapid to allow for any significant inflation.

The second issue is one of detectability. Current state-of-the-art transit surveys like NASA's {\it Kepler} mission can detect planets with transit depths down to a few hundred parts per million \cp{Borucki2010}. For inflated hot Jupiters this detection limit means that planets could only be found around stars with radii $\lesssim$ 10 R$_{\mathrm{\odot}}$. For a solar-mass star, this corresponds to stars at the bottom of the red giant branch, just past the red bump and before the start of helium core burning \cp{KippenhahnBook}. Is it possible to produce significantly inflated radii around stars that are still small enough that the inflated planets will be detectable?

Fortunately, as we will show in Section \ref{resultssec} there is a range of parameter space in which it is possible to produce inflated planets around evolved stars that are detectable with current surveys. In particular, planets with orbital periods of $\sim$ 10-30 days are ideal candidates to search for the signature of inflationary heating around evolved stars. Around main sequence stars, giant planets in this period range are warm Jupiters with incident fluxes $\lesssim$100 F$_{\mathrm{\oplus}}$ and temperatures $\lesssim$1000 K. Such planets are cool enough that there should be no inflated planets around main sequence stars \cp{Miller2011,Demory2011} as shown in Figure \ref{inffig}, making their post-main sequence evolution a clear test of inflation models. However, planets in this period range are still irradiated enough that it is possible to significantly inflate their radii before their parent star reaches 10 R$_{\mathrm{\odot}}$ if at least $\sim$1\% of their incident irradiation is deposited into their deep interior. We term these planets "re-inflated" warm Jupiters.

The outline of this paper is as follows: in Section \ref{modelsec} we will discuss our stellar and planetary evolution models. Then in Section \ref{resultssec} we will examine the parameter space that allows for re-inflated planets. Finally, in Section \ref{surveysec} we will discuss the prospects for detecting re-inflated planets, if they exist, with current and upcoming transit surveys.


\section{Evolution Models}
\label{modelsec}

In order to understand the circumstances that can lead to re-inflated radii, we need a thermal evolution model for giant planet interiors, which can predict planet radii for planets with different masses, ages, levels of irradiation, and inflationary heating mechanisms. Here we use a modified version of the planet interior and thermal evolution model described in \ct{Lopez2012} and \ct{Lopez2014}. This model has previously been used to predict planetary radii across parameter space for low-mass planets \cp{Lopez2014} and to examine the impact of photo-evaporation such planets \cp{Lopez2012, Lopez2013}. For further details, we refer the reader to Sections 2 \& 3 of \ct{Lopez2014}, however, we will briefly summarize the models here.

At any point along its evolutionary track, a planet model is defined by the total mass, the amount of incident radiation it receives, and the internal specific entropy of its H/He envelope. As in our previous work, we use the solar-composition \ct{Baraffe1998} stellar evolution tracks to calculate stellar luminosities for the pre-main sequence and main sequence evolution of the host stars. In addition, here we also include post-main sequence evolution from the Padova stellar evolution models, again assuming solar composition \cp{Bertelli2008}. 

Throughout this paper, we assume a 10 M$_{\mathrm{\oplus}}$ Earth-like rock \& iron core with an Earth-like 2:1 rock/iron ratio, using the ANEOS olivine \ct{Thompson1990} and SESAME 2140 Fe \ct{Lyon1992} equations of state (EOS). Generally, however, the presence of this core does not contribute significantly to our predicted planet radii. For the H/He envelope we assume solar composition and a fully adiabatic interior using the \ct{Saumon1995} EOS.  Finally, atop the H/He envelope is a thin radiative atmosphere, which we assume is isothermal at the planet's equilibrium temperature. We define a planet's radius at 20 mbar, appropriate for a transit viewing geometry \cp{Hubbard2001}.

These individual structure models are then connected in time by tracking the net heating and cooling of a planet's interior, and evolving the entropy of the H/He adiabat accordingly. In order to track radiative cooling from the planetary atmosphere, we use a grid of solar metallicity atmosphere models computed over a range of surface gravities and incident fluxes from \ct{Fortney2007}. These atmosphere models are fully non-gray, i.e., wavelength dependent radiative transfer is performed rather than simply assuming a single infrared opacity. In addition, our model by default tracks heating due radiogenic decay in the planet's rocky core, along with the slight delay in the cooling of the envelope due to the need to cool the rock/iron core as well. However, for the giant planets that we consider here, these terms make only a minor contribution to a planet's overall energy budget.

As with previous work, we assume that planets initially form with a large initial entropy according to the traditional "Hot-Start" model \cp[e.g.,][]{Fortney2007, Marley2007}. However, since we are examining the radii of planets around post-main sequence stars, our results are completely insensitive to this choice of initial condition. Unlike \ct{Lopez2012} or \ct{Lopez2013}, however, we do not consider the effects of photo-evaporative mass loss on planet evolution. This is unlikely to effect our results, since we are considering giant planets at relatively modest levels of irradiation, at least while their host stars are on the main sequence, so they are unlikely to experience any significant evaporation. In total, our model is described by equation (\ref{thermaleq}).

\begin{equation}\label{thermaleq}
\int_{M_{\mathrm{core}}}^{M_{\mathrm{p}}} dm \frac{T dS}{dt} = - L_{\mathrm{int}} + L_{\mathrm{radio}} - c_{\mathrm{v}} M_{\mathrm{core}}  \frac{dT_{\mathrm{core}}}{dt}+\epsilon_{\mathrm{heat}} F_{\mathrm{p}}\pi R_{\mathrm{p}}^2
\end{equation}

Here the left hand side describes the cooling rate of a planet's H/He adiabat and the right hand side the various heating and cooling terms. $L_{\mathrm{int}}$ describes the atmospheric cooling rate according to the \ct{Fortney2007} models, $L_{\mathrm{radio}}$ describes radiogenic heating in the rocky core, and $c_{\mathrm{v}} M_{\mathrm{core}}  \frac{dT_{\mathrm{core}}}{dt}$ describes the delay in cooling of the envelope due to the need to cool the core as well. Here, we make one simple modification to the thermal evolution model described in \ct{Lopez2014}, and that is the introduction of additional heating term to account for a generalized inflationary heating mechanism. This additional term is defined by a heating efficiency $\epsilon_{\mathrm{heat}}$, which defines the fraction of a planet's incident stellar irradiation $F_{\mathrm{p}}$ which is mixed into the planet's interior adiabat. 

\begin{figure}
  \begin{center}
    \includegraphics[width=3.5in,height=2.5in]{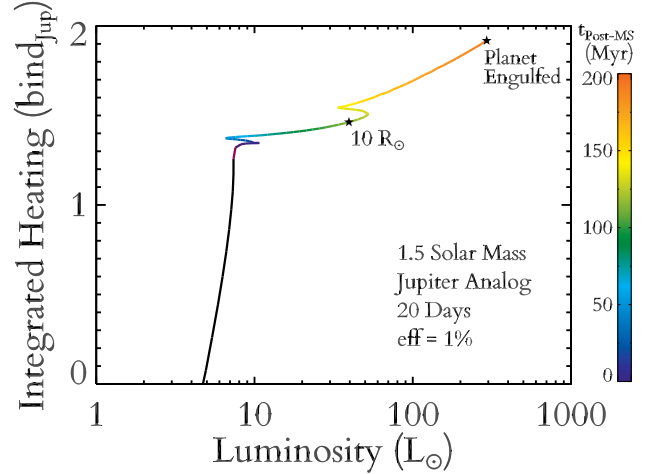}
  \end{center}
  \caption{Here we show the integrated inflationary heating that a planet receives as its parent star evolves off the main sequence for a Jupiter mass planet with a 20 day orbit around a 1.5 M$_{\mathrm{\odot}}$ star. The x-axis shows the evolution of the stellar luminosity according to Padova stellar evolution tracks \cp{Bertelli2008}, while the y-axis shows the cumulative heating that our example planet receives, assuming a 1\% heating efficiency, compared to the current binding energy of Jupiter. The black portion of the curve shows the main-sequence evolution, while color coding indicates the time since the parent star left the main sequence in Myr. Black stars indicate when the stellar radius reaches 10 $R_{\mathrm{\odot}}$, the normal stopping point for our models, and when the star engulfs the planet's orbit. \label{rgbfig}}
\end{figure}

Figure \ref{rgbfig} illustrates some of the key features of our model and provides a simple illustration for the potential of moderate period giant planets to re-inflate as their parent stars move up the red giant branch. Here we show the cumulative energy deposited into a planetary interior by the extra inflationary heating term in equation (\ref{thermaleq}), as its host star moves up the sub-giant and red-giant branches. This example is for Jupiter mass planet with a 20 day orbit around a 1.5 M$_{\mathrm{\odot}}$ star, with a interior heating efficiency of 1\%, typical of the models that we will describe in Section \ref{resultssec}. During the parent star's post main-sequence evolution, the heating term deposits $\sim$half the binding energy of Jupiter into our example planet's interior in just 150 Myr.

In addition to our parameterized heating model, it is of course also valuable to examine the results of various detailed models of inflationary heating. Implementing a full inflationary heating model in our planet evolution code is beyond the scope of this paper.  However, various authors provide analytic approximations for the inflationary heating rates predicted by their models. Here we select two such models which we have implemented into our evolution code in addition to the parameterized heating efficiency described in equation (\ref{thermaleq}).

Equation (\ref{Huangeq}) is taken from the ohmic dissipation models of \ct{Huang2012}, equation (14) of that paper. Their models provide a good example of an inflationary model that stalls but generally cannot reverse planetary contraction. Equation (\ref{Huangeq}) gives the ohmic heating rate as function of the planetary mass, radius, and equilibrium temperature, along with $B_{\mathrm{\phi,0}}$ the toroidal component of the planetary magnetic field and $\sigma_{\mathrm{t}}$ the electric conductivity in the dissipation region. For our models here we leave these final two parameters fixed at the nominal values given by \ct{Huang2012}.

\begin{equation}\label{Huangeq} 
\begin{split}
L_{\mathrm{Huang}} \approx 3 \times 10^{22} & \mathrm{erg \, s^{-1}} \left(\frac{B_{\mathrm{\phi,0}}} {10 \,\mathrm{G}} \right)^2 \left(\frac{\sigma_{\mathrm{t}}}{10^6 \, \mathrm{s}^{-1}}\right)^{-1} \\
 & \times \left(\frac{T_{\mathrm{eq}}}{1500 \, \mathrm{K}}\right) \left( \frac{R_{\mathrm{P}} }{R_{\mathrm{J}}} \right)^{4} \left(\frac{M_{\mathrm{p}}}{M_{\mathrm{J}}}\right)^{-1}
\end{split}
\end{equation}

Equation (\ref{Socrateseq}) meanwhile, provides the approximate maximum heating rate from thermal tides according to \ct{Socrates2013}, equation (8) of that paper. 

\begin{equation}\label{Socrateseq}
\begin{split}
L_{\mathrm{Socrates}} =  1.5 \, \times & \, 10^{28}  \mathrm{\, erg \, s^{-1} \,} \left(\frac{P}{4 \, \mathrm{days}}\right)^{-2} \\
 & \times \left(\frac{T_{\mathrm{eq}} }{2000 \, \mathrm{K}}\right)^{3} \left(\frac{R_{\mathrm{P}}}{10^{10} \, \mathrm{cm}}\right)^{4} 
\end{split}
\end{equation}

Although both models describe heating rates that goes as $\propto R_{\mathrm{P}}^{4}$, comparing equations (\ref{Huangeq}) and (\ref{Socrateseq}) shows that the thermal tide models predict a much stronger dependence on planetary irradiation. While the models of \ct{Huang2012} are only linear in planetary temperature, the thermal tide models of \ct{Socrates2013} go as $\propto T_{\mathrm{eq}}^{3}$, in addition to the separate period dependence due to the planet's gravitational tides. As a result, we should expect that the models of \ct{Socrates2013} should be much more amenable to producing re-inflated planets than those of \ct{Huang2012}.

Finally, there is one other potential ingredient that we may wish to include in our models. Orbital decay due to tidal evolution depends strongly on the radius of the host star, therefore it is reasonable to ask whether including tidal evolution along with the stellar, and planetary thermal evolution will significantly alter our results. To first order in eccentricity the orbital decay timescale for a planet is given by \cp{Mardling2002}

\begin{equation}
\frac{1}{\tau_{\mathrm{a}} } = \left( \frac{ \dot{a}_{\mathrm{p}} }{ a_{\mathrm{p}} }\right )_{\mathrm{star}} +\left(\frac{ \dot{a}_{\mathrm{p}} }{ a_{\mathrm{p}} }\right)_{\mathrm{planet}}
\end{equation}

\noindent where

\begin{equation}
\left(\frac{ \dot{a}_{\mathrm{p}} }{a_{\mathrm{p}} }\right)_{\mathrm{star}} = - \frac{9}{2} \left(\frac{ n_{\mathrm{p}} }{ Q_{\mathrm{s}}^\prime  }\right) \left(\frac{ M_{\mathrm{p}} }{ M_{\mathrm{s}}  }\right) \left(\frac{ R_{\mathrm{s}} }{ a_{\mathrm{p}} }\right)^{5} \left[ 1 - \left(\frac{ P_{\mathrm{orb}} }{ P_{\mathrm{spin}} }\right) \right]
\end{equation}

\noindent is the term due to tides within the star and

\begin{equation}
\left(\frac{ \dot{a}_{\mathrm{p}} }{a_{\mathrm{p}} }\right)_{\mathrm{planet}} = - \frac{171}{4} \left(\frac{ n_{\mathrm{p}} }{ Q_{\mathrm{p}}^\prime  }\right) \left(\frac{ M_{\mathrm{s}} }{ M_{\mathrm{p}} }\right) \left(\frac{ R_{\mathrm{p}} }{ a_{\mathrm{p}} }\right)^{5} e_{\mathrm{p}}^{2}
\end{equation}

\noindent is the term due to tides within the planet. Here ${a}_{\mathrm{p}}$ is the planet's semi-major axis; ${n}_{\mathrm{p}}$ is its orbital frequency; ${Q}_{\mathrm{p,s}}^\prime$, ${M}_{\mathrm{p,s}}$, and ${R}_{\mathrm{p,s}}$ are the modified tidal Q values, Masses, and Radii for the planet and host star, respectively; ${e}_{\mathrm{p}}$ is the planet's eccentricity; and ${P}_{\mathrm{orb}}$ and ${P}_{\mathrm{spin}}$ are the orbital period and stellar spin period. Let us consider a 1 ${M}_{\mathrm{J}}$, 1.5 ${R}_{\mathrm{J}}$ planet at 10 days orbiting a 1 ${M}_{\mathrm{\odot}}$, 10 ${R}_{\mathrm{\odot}}$ star, typical values for the parameter space that we are interested in. Assuming ${Q}_{\mathrm{p}}^\prime$ of 10$^5$, ${Q}_{\mathrm{s}}^\prime$ of 10$^7$, and a reasonable eccentricity then the stellar tidal term will dominate over the planet term. Likewise, for a 10 ${R}_{\mathrm{\odot}}$ star ${P}_{\mathrm{spin}}$ will be $\gg {P}_{\mathrm{orb}}$ , allowing us to simplify further. Plugging in our example values gives a tidal decay timescale of 290 Myr, which while short compared to the main sequence lifetime, is quite long compared to the stellar evolution timescale for a star at the bottom of the red giant branch to evolve until it is $\gg$10 ${R}_{\mathrm{\odot}}$. As a check, we reran all of the models presented in Section \ref{resultssec} with tidal evolution and an initial planetary eccentricity of 0.1. Including these terms in our models, had no noticeable effect on our results.

\section{Results}
\label{resultssec}

\begin{figure}
  \begin{center}
    \includegraphics[width=3.3in,height=4.9in]{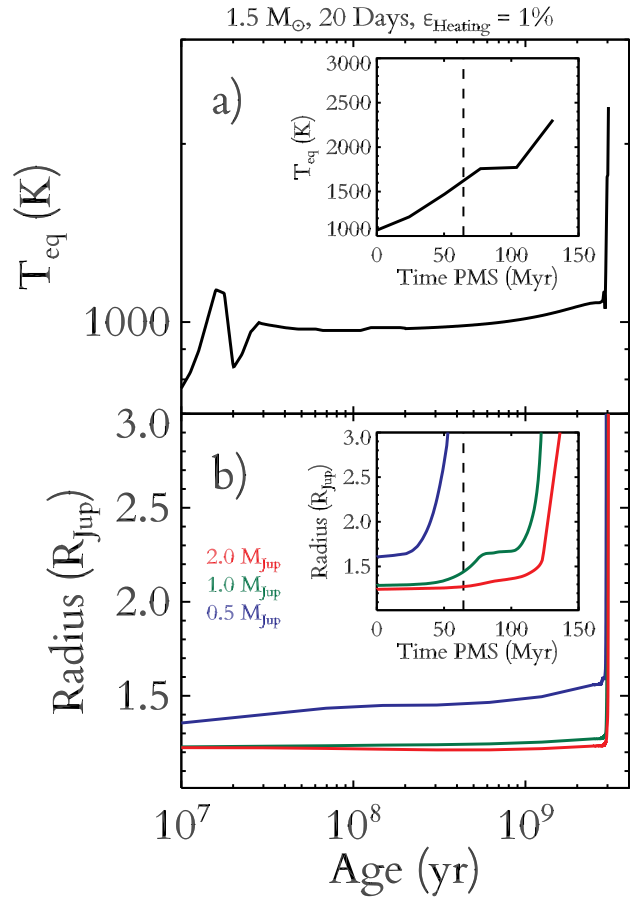}
  \end{center}
  \caption{Three example thermal evolution models for warm Jupiters with a simple prescription for re-inflation. As a representative example we chose a 20 day orbit around a 1.5 M$_{\mathrm{\odot}}$ star. In panel a) we show the evolution of planetary equilibrium temperature as the star evolves along the main sequence and up the red giant branch. In panel b) we show the corresponding radius evolution for planets that are 0.5, 1.0, and 2.0 Jupiter masses, assuming that at any given age 1\% of the incident irradiation is deposited into the deep interior. In each case the insets show the last 150 Myr, starting when the star leaves the main sequence. The vertical dashed lines in these insets indicate when the star reaches 10 R$_{\mathrm{\odot}}$, the normal stopping point in our full grid of models. During this time the star's luminosity increases rapidly and the incident flux at 20 days becomes comparable to that on traditional hot Jupiters. Lower mass planets are easier to re-inflate; planets with a mass similar to Saturn rapidly inflate to over 2 Jupiter radii. \label{exfig}}
\end{figure}

\begin{figure}
  \begin{center}
    \includegraphics[width=3.5in,height=2.5in]{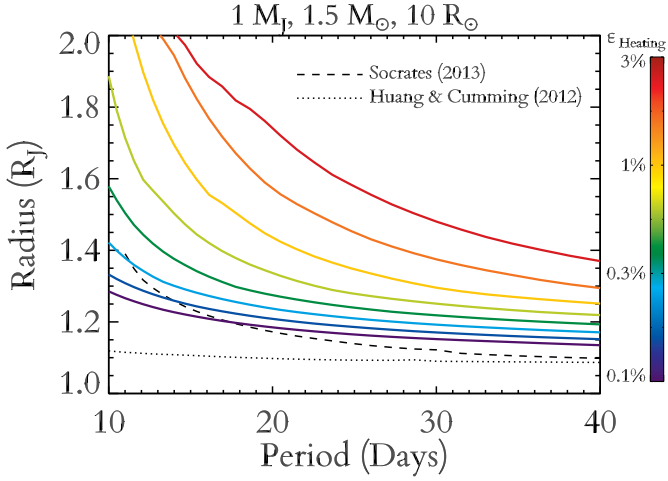}
  \end{center}
  \caption{Here we show re-inflated planet radii in Jupiter radii vs. orbital period in days for different heating efficiencies and inflation models. All models assume a Jupiter mass planet orbiting a 1.5 M$_{\mathrm{\odot}}$ star that has evolved up the red giant branch until it reached 10 R$_{\mathrm{\odot}}$. The dotted line shows an analytic approximation to the ohmic heating model of \ct{Huang2012} as described by equation (\ref{Huangeq}), while the dashed line shows the thermal tide model of \ct{Socrates2013} as described by equation{Socrateseq}. \label{periodfig}}
\end{figure}

\begin{figure}
  \begin{center}
    \includegraphics[width=3.5in,height=2.5in]{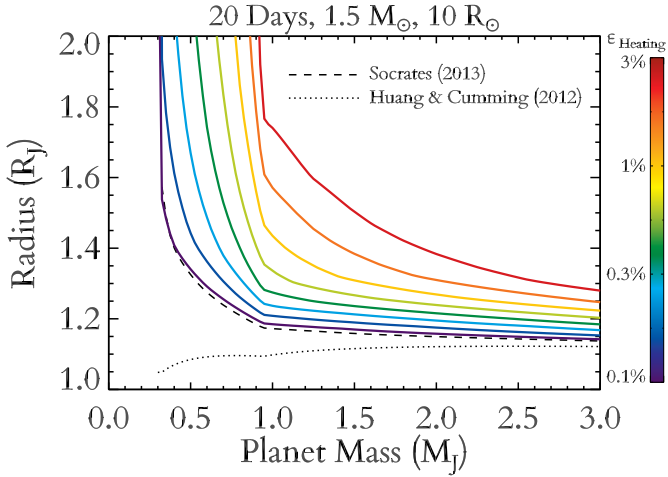}
  \end{center}
  \caption{ Similar to Figure \ref{periodfig}, except here we fix the orbital period at 20 days and vary the planet mass. \label{massfig}}
\end{figure}

\begin{figure}
  \begin{center}
    \includegraphics[width=3.5in,height=2.5in]{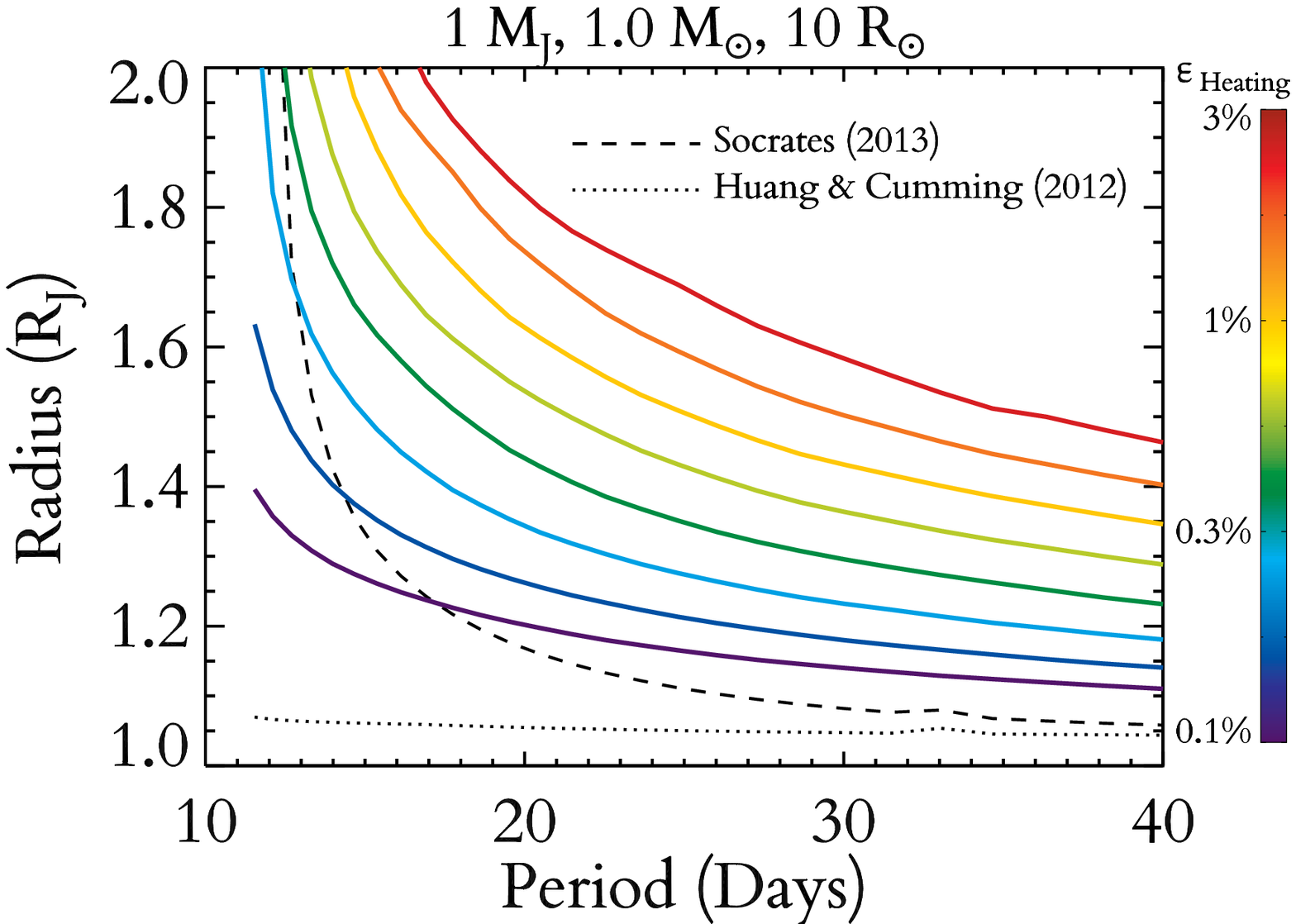}
  \end{center}
  \caption{ Similar to Figure \ref{periodfig}, but with a 1.0  M$_{\mathrm{\odot}}$ star instead. \label{periodsolarfig}}
\end{figure}

\begin{figure}
  \begin{center}
    \includegraphics[width=3.5in,height=2.5in]{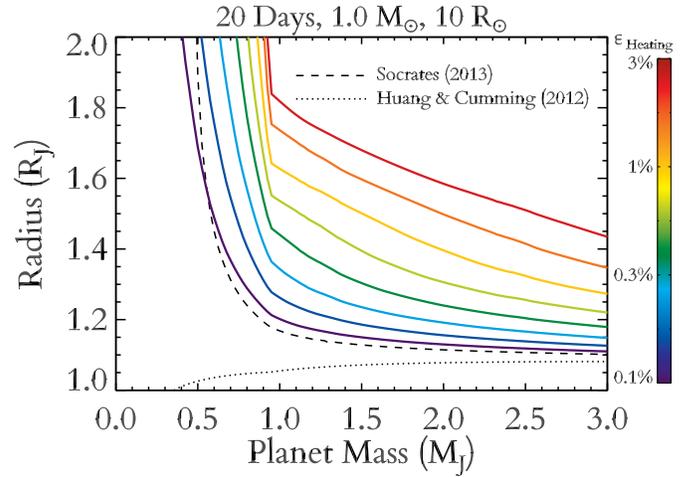}
  \end{center}
  \caption{  Similar to Figure \ref{massfig}, but with a 1.0  M$_{\mathrm{\odot}}$ star instead. \label{masssolarfig}}
\end{figure}

With this relatively straight-forward evolution model, we can now begin to explore the circumstances that can allow for the re-inflation of giant planets around evolve stars. To this end we ran a grid of planet evolution models over a range of planet masses, orbital periods, and interior heating efficiencies. In each case, we placed the planet around a 1.5 M$_{\mathrm{\odot}}$ star and tracked the planet's thermal evolution until the host star reached 10 R$_{\mathrm{\odot}}$. The stellar mass of 1.5 M$_{\mathrm{\odot}}$ was chosen since this value is representative of the giant stars with known KOIs in the \ct{Huber2013} asteroseismology study. Our general conclusions are insensitive to this choice, beyond the fact that,  all else being equal, planets around slightly less massive stars will experience slightly more re-inflation since their host stars move more slowly up the sub-giant and red-giant branches. For our study we varied the planet mass from 0.3 to 3 M$_{\mathrm{J}}$, the orbital period from 10 to 40 days, and the interior heating efficiency $\epsilon_{\mathrm{heat}}$  from 0.1\% to 3\%. We chose this range of planet masses in order to cover the range of inflated hot Jupiters orbiting main sequence stars seen in Figure \ref{mrfig}. Likewise, our range of heating efficiencies is meant to bracket the range of efficiencies predicted by models trying to explain the normal inflated planet population. However, as mentioned above we have deliberately chosen a longer orbital period range in order to exclude planets that could have been inflated while in the main sequence, that we are only testing whether planets can become re-inflated during their post-main sequence lifetimes. Lastly, we also ran models over our planet mass and period grid for each of the three semi-analytic models described in Section \ref{modelsec}.

Figure \ref{exfig} shows example evolution runs for three different planet masses orbiting at 20 days around a 1.5 $M_{\mathrm{\odot}}$ star. Here we show the full radius evolution path for each planet, assuming a heating efficiency of 1\% and continuing the evolution until the planets are swallowed by their host star. Although, it is quite difficult to re-inflate the 1 or 2 M$_{\mathrm{J}}$ planets before the star reaches 10 R$_{\mathrm{\odot}}$, re-inflation is very plausible for the 0.5 planet M$_{\mathrm{J}}$ model.

Figures \ref{periodfig} and \ref{massfig}, meanwhile, show the results of our full parameter study. In Figure \ref{periodfig} we fix the planet's mass at 1 M$_{\mathrm{J}}$ and explore how the final planetary radius predicted by our models varies with orbital period for different values of our parameterized heating efficiency, along with the analytic approximation to the models of \ct{Huang2012} and \ct{Socrates2013} as described in equations (\ref{Huangeq}) and (\ref{Socrateseq}). Similarly, in Figure \ref{massfig} we fix the orbital period at 20 days and instead vary the mass or our model planets.

Examining Figures \ref{periodfig} and \ref{massfig}, we see that re-inflated planets, if they exist, are most likely to occur for planets with masses $\lesssim$1 M$_{\mathrm{J}}$ and orbital periods $\lesssim$15 days. For such planets our models predict that planets should have radii that are clearly inflated and detectable for heating efficiencies >0.1\%. Likewise, in this region we see a clear difference between the two analytic models described in Section \ref{modelsec}. While the \ct{Huang2012} formula does not lead to re-inflation in any of the models we ran, the \ct{Socrates2013} prescription produces significant re-inflation at low planet masses and short orbital periods. Furthermore, as shown in Figures \ref{periodsolarfig} and \ref{masssolarfig}, the mass and period range allowing for re-inflation will be larger for slightly less massive stars since these spend significantly more time on the sub-giant and red-giant branches allowing for more heating. Likewise, as shown in Figure \ref{exfig} re-inflation becomes much easier around stars larger than 10 R$_{\mathrm{\odot}}$, although as discussed above planets around larger stars will be difficult to detect. This creates an easily searchable region of planet parameter space where we can test for the existence of re-inflation planets and thereby constrain models of hot Jupiter inflation.

Figures \ref{periodfig}-\ref{masssolarfig}, also raise another intriguing possibility for the detection of re-inflated planets. In some cases, particularly for models with masses <1 M$_{\mathrm{J}}$, our models lead to planet with radii > 2 R$_{\mathrm{J}}$. If possible, such large radii would be an even clearer sign of re-inflation, as this is larger than any currently known hot-Jupiters. Moreover, at an orbital period of 20 days around a Sun-like star the hill radius of a Jupiter mass planet is $\sim$ 20 R$_{\mathrm{J}}$, meaning that unlike hot Jupiters such planets would not be vulnerable to Roche-lobe overflow. It is unclear, however, whether normal inflationary heating mechanisms can continue to operate in such large planets as they not been tested in any of the models of normal hot Jupiter inflation. Furthermore, scaling laws for hot-Jupiter circulation suggest that atmospheric wind speeds should decrease with increasing planet radius, indicating that ohmic dissipation, at least, is likely to be ineffective in this regime (see Equation (13) in \ct{Showman2002}). Nonetheless, though the possibility of re-inflated planets with radii > 2 R$_{\mathrm{J}}$ is an intriguing possibility both for transit surveys and for models of hot-Jupiter inflation.

\section{Searching for Re-inflated Planets}
\label{surveysec}

From Figures \ref{periodfig} and \ref{massfig} we can see the ideal parameter space to test whether or not re-inflated planets exist, and therefore the nature of planet inflation. Re-inflated planets are most likely to both exist and be detectable at periods between 10 and 20 days around post main sequence stars that have moved up the sub-giant branch and have stellar radii $\sim$5-10 R$_{\mathrm{\odot}}$. Planets are easiest to re-inflate if they have masses $\lesssim$1 M$_{\mathrm{J}}$, however, planet mass is of course something not known a priori. 

Ideally any search for re-inflated planets would be complete for planets down to slightly less than 1 R$_{\mathrm{J}}$ (corresponding to transit depths of $\sim$100 ppm around a 10 R$_{\mathrm{\odot}}$ star) and include a range of periods bracketing this ideal 10-20 day range in order to test the null hypothesis as well. Finally it is important that any planet candidates have reliable stellar radii determinations that are accurate to at least $\sim$ 10\%, in order to reliably distinguish between inflated and non-inflated planets. Fortunately, this accuracy is readily achievable since evolved stars have large amplitude asteroseismic modes that make it relatively easy to determine asteroseismic stellar radii \cp[e.g.,][]{Huber2013}.

The transit depth of an inflated giant planet around a 10 R$_{\mathrm{\odot}}$ star is comparable to the depth of a super-Earth sized planet around a sun-like star, a precision routinely achieved by NASA's incredibly successful {\it Kepler} mission. With over 16 quarters of data and a significant sub-sample of post main-sequence stars, it is worth asking if there are potentially re-inflated giant planets currently in the literature. Stellar parameters from the Kepler Input Catalog are notoriously unreliable for late spectral types, so it is wise to limit our search to Kepler host stars with asteroseismically determined stellar radii. \ct{Huber2013} lists 107 Kepler Objects of Interest or KOIs orbiting 77 host stars with asteroseismic stellar parameters. Of that list only five KOIs orbit stars with stellar radii >5 R$_{\mathrm{\odot}}$. Of these five, one (KOI 981.01) is a hot-Neptune and two KOIs (1230.01 and 2481.01) are clear asteroseismic false positives.

Initially, however, KOI 2640.01 looks very promising. It has an orbital period of 33 days around a 1.27$\pm$0.11 M$_{\mathrm{\odot}}$ retired F-star that is 7.48$\pm$0.25 R$_{\mathrm{\odot}}$. This puts it close to the range where we would re-inflated planets to exist. Moreover with a planetary radius of 1.52$\pm$0.07 R$_{\mathrm{J}}$, the candidate, if real, would be unambiguously inflated. Unfortunately, further examination suggests that KOI 2640.01 is likely also a false positive. Recently, \ct{Sliski2014} examined the asteroseismic stellar densities of giant stars in the \ct{Huber2013} sample and compared those to the mean stellar densities predicted by the KOIs' orbits and transit durations. \ct{Sliski2014} found that the two stellar densities were completely inconsistent for KOI 2640.01, implying that the candidate is likely a false positive. Moreover, an inspection of the individual transits for this object (Daniel Huber, private communication), shows that the apparent depth varies significantly from transit to transit, suggesting that the signal is due to stellar granulation noise.

Finally, we are left with KOI 2133.01, also known as Kepler-91b, which was confirmed to be a planet by \ct{Lillo-Box2014a}. Subsequent radial velocity observations in \ct{Lillo-Box2014b} and \ct{Barclay2014} have re-affirmed its planetary nature and found that it has a mass of 0.76$\pm$0.13 M$_{\mathrm{J}}$ \cp{Barclay2014}. With a radius of 1.31$\pm^{0.06}_{0.07}$ R$_{\mathrm{J}}$, Kepler-91b is definitely inflated. Moreover, at 1.3 M$_{\mathrm{\odot}}$ and 6.3 R$_{\mathrm{\odot}}$ its host star Kepler-91 is a red giant, making Kepler-91b appear like a prime target for re-inflation. However, with an orbital period of just 6.2 days Kepler-91b is sufficiently close-in that it could easily have been inflated even while its host star was still on the main sequence. When Kepler-91 was still on the main sequence, 91b received $\sim$1300 F$_{\mathrm{\oplus}}$. Its current inflated size is typical of the known inflated planets shown in Figure \ref{mrfig} that receive this level of irradiation. As a result, we cannot rule out the possibility that Kepler-91b was already inflated before its star moved up the red giant branch, making it an ineffective test of re-inflation.

In summary then, there are currently no {\it Kepler} candidates that are possible planets for re-inflation. Despite this lack of current targets, there is plenty of reason to hope that current and upcoming surveys will be able to search for this possible new population. Although the original {\it Kepler} target sample \cp{Brown2011} did include a small subset of giant stars for asteroseismology \cp[e.g.,][]{Huber2010}, few of these giants are large enough to allow for re-inflation while still being small enough to allow for transiting planets to be detected. Currently, however the {\it Kepler} is in midst of the new {\it K2} survey of multiple 100 sq degree fields in the ecliptic plane \cp{Howell2014}. Each $\sim$80 day {\it K2} campaign allows for new targets with different selection criteria. This provides an ideal opportunity to search for a sample of giant planets at our ideal period range of 10-30 days around $\sim$10 R$_{\mathrm{\odot}}$ stars. Moreover, looking beyond the {\it Kepler} era, in 2017 NASA will launch the {\it TESS} mission \cp{Ricker2014} and in 2024 ESA will launch {\it PLATO} \cp{Rauer2014}, both of which will have sufficient depth and period coverage to find re-inflated planets as well. If and when any re-inflated planets are found, it should be relatively easier to get radial velocity mass determinations for them given their large masses and moderate orbital periods. With both masses and radii for such planets will be able to determine the over-all energy budget for such planets and therefore constrain the efficiency of inflationary heating with the models shown in Figures \ref{periodfig}-\ref{masssolarfig}.

\section{Summary}

The unknown heating mechanism that produces large inflated radii among the most irradiated hot Jupiters is one of the great outstanding challenges for the theory of extrasolar giant planets. By examining the post-main sequence lives of moderate period "warm Jupiters", it may be possible to finally distinguish between the different possible explanations of hot Jupiter inflation. If the heating mechanism operates in a manner that deposits some fraction of a planet's incident irradiation into its deep interior, then as its host star moves up the sub-giant and red-giant branches, the planet's interior should heat up and its radius should increase in response. In contrast, if the inflation mechanism operates by simply slowing radiative cooling and contraction, thereby allowing a planet to retain more of its initial heat from formation, then there should be no increase in planetary radius as the irradiation increases. For planets with masses <1 M$_{\mathrm{J}}$ and orbital periods of $\sim$10-20 days around 1-1.5 M$_{\mathrm{\odot}}$ stars, re-inflation should occur for even modest interior heating efficiencies. This creates a highly detectable region of parameter space that will allow a unique test of the physics of hot Jupiter inflation. Although we currently lack targets that are well suited to testing re-inflation, there is strong reason to hope that upcoming surveys will be able to search for this important potential population.

\acknowledgements{We would like to thank Daniel Huber, Mike Endl, Yann Alibert, and Angie Wolfgang for their input and many helpful discussions. This research has made use of the Exoplanet Orbit Database and the Exoplanet Data Explorer at exoplanets.org." We acknowledge the support of NASA grant NNX09AC22G and NSF grant AST-1010017. The research leading to these results also received funding from the European Union Seventh Framework Programme (FP7/2007-2013) under grant agreement number 313014 (ETAEARTH).} 

\bibliographystyle{apj}
\bibliography{myreferences}
\
\end{document}